\documentclass[twocolumn,final]{svjour3}
\usepackage{multicol}
\usepackage[utf8]{inputenc}
\usepackage{caption}
\usepackage{times,amsfonts,graphicx}
\usepackage[perpage,symbol]{footmisc}
\usepackage[english]{babel}
\usepackage{multirow,multicol}
\usepackage{amsmath} 
\usepackage{pbox, booktabs, rotating}
\usepackage{subfigure}
\usepackage{array, makecell, lipsum}
\usepackage{enumerate}
\usepackage{threeparttable}
\usepackage{multirow,hyphenat,balance}
\usepackage{placeins}
\usepackage[compress]{cite}
\usepackage{float,color}
\usepackage{ragged2e}
\usepackage{blindtext}
\usepackage{enumerate}
\usepackage{etoolbox}
\usepackage{gensymb}
\usepackage{rotating, color} 
\usepackage[linesnumbered,ruled]{algorithm2e}
\usepackage{url}
\usepackage{textcomp}
\usepackage{soul,xcolor}
\usepackage[switch,pagewise]{lineno}
\usepackage{soul,xcolor}
\setstcolor{red}
\usepackage{flushend} 
\usepackage{lipsum}

\begin{document} \sloppy

\newcolumntype{M}[1]{>{\centering\arraybackslash}m{#1}}
\newcommand{\algrule}[1][.2pt]{\par\vskip.5\baselineskip\hrule height #1\par\vskip.5\baselineskip}
\graphicspath{{./figs/}}
\def\todo#1{\textbf{\color{red}[TODO:~#1]}}
\def\todoYuqiao#1{\textbf{\color{red}[Yuqiao:~#1]}}
\def\note#1{\textbf{\textcolor{blue}{[#1]}}}
\def\update#1{\textbf{\textcolor{red}{[Need to update #1]}}}

\newcommand*{\ie}{i.e.,\@\xspace}
\newcommand*{\eg}{e.g.,\@\xspace}
\newcommand*{\el}{et al.\@\xspace}

\title{A Novel Topology-Guided Attack and Its Countermeasure Towards Secure Logic Locking}

\author{Yuqiao Zhang
\and Ayush Jain \and Pinchen Cui \and Ziqi Zhou \and Ujjwal Guin}

\institute{
Y. Zhang \and A. Jain \and Z. Zhou \and U. Guin
\at Department of Electrical and Computer Engineering, \\Auburn University, Auburn, AL, USA
\\\email{\\\{yuqiao.zhang, ayush.jain, ziqi.zhou, ujjwal.guin\}@auburn.edu}
\and
P. Cui
\at Department of Computer Science and Software Engineering, \\Auburn University, Auburn, AL, USA\\
\email{pinchen@auburn.edu}
}




\maketitle
 

\begin{abstract}
The outsourcing of the design and manufacturing of integrated circuits~(ICs) in the current horizontal semiconductor integration flow has posed various security threats due to the presence of untrusted entities, such as overproduction of ICs, sale of out-of-specification/rejected ICs, and piracy of Intellectual Properties (IPs). Consequently, logic locking emerged as one of the prominent design for trust techniques. Unfortunately, these locking techniques are now inclined to achieve complete Boolean satisfiability (SAT) resiliency after the seminal work published in \cite{SubramanyanHost2015}. In this paper, we propose a novel oracle-less attack that is based on the topological analysis of the locked netlist even though it is SAT-resilient. The attack relies on identifying and constructing unit functions with a hypothesis key to be searched in the entire netlist to find its replica. The proposed graph search algorithm efficiently finds the duplicate functions in the netlist, making it a self-referencing attack. This proposed attack is extremely efficient and can determine the secret key within a few minutes. We have also proposed a countermeasure to make the circuit resilient against this topology-guided attack to progress towards a secure logic locking technique. 

\vspace{-5px}
\keywords{
Logic locking \and Boolean satisfiability \and Boolean functions \and piracy \and overproduction \and directed graph \and depth-first search}
\end{abstract}


\section{Introduction}
The prohibitive cost of building and maintaining a foundry~(fab) with advanced technology nodes has forced many design companies to become fabless and adopt the horizontal semiconductor integration model. Currently, majority of the design houses integrates intellectual properties~(IPs) obtained from different third-party IP~(3PIP) vendors along with its design and outsources the manufacturing to an offshore foundry resulting in a global supply chain with distributed vendors carrying out design, verification, fabrication, testing, and distribution of chips. The involvement of untrusted entities at various stages in the IC manufacturing and testing process has resulted in evident security threats, such as piracy or theft of IPs, overproduction of ICs, and sale of out-of-specification/rejected ICs~\cite{Alkabani2007, Castillo2007TVLSI, GuinTODAES2016, Chakraborty4681649, bhunia2018hardware, GuinSpringer2014}. Many design-for-trust techniques have been studied over the years as countermeasures against the aforementioned threats~\cite{Roy4484823, RajendranDAC2012, GuinTODAES2016, Koushanfar01DAC, Alkabani2007, CharbonCICC98, KahngICCAD2001, qu2003intellectual, jarvis2007split, vaidyanathan2014efficient}. 

\begin{figure}[t]
\centering
\includegraphics[width=1.0\linewidth]{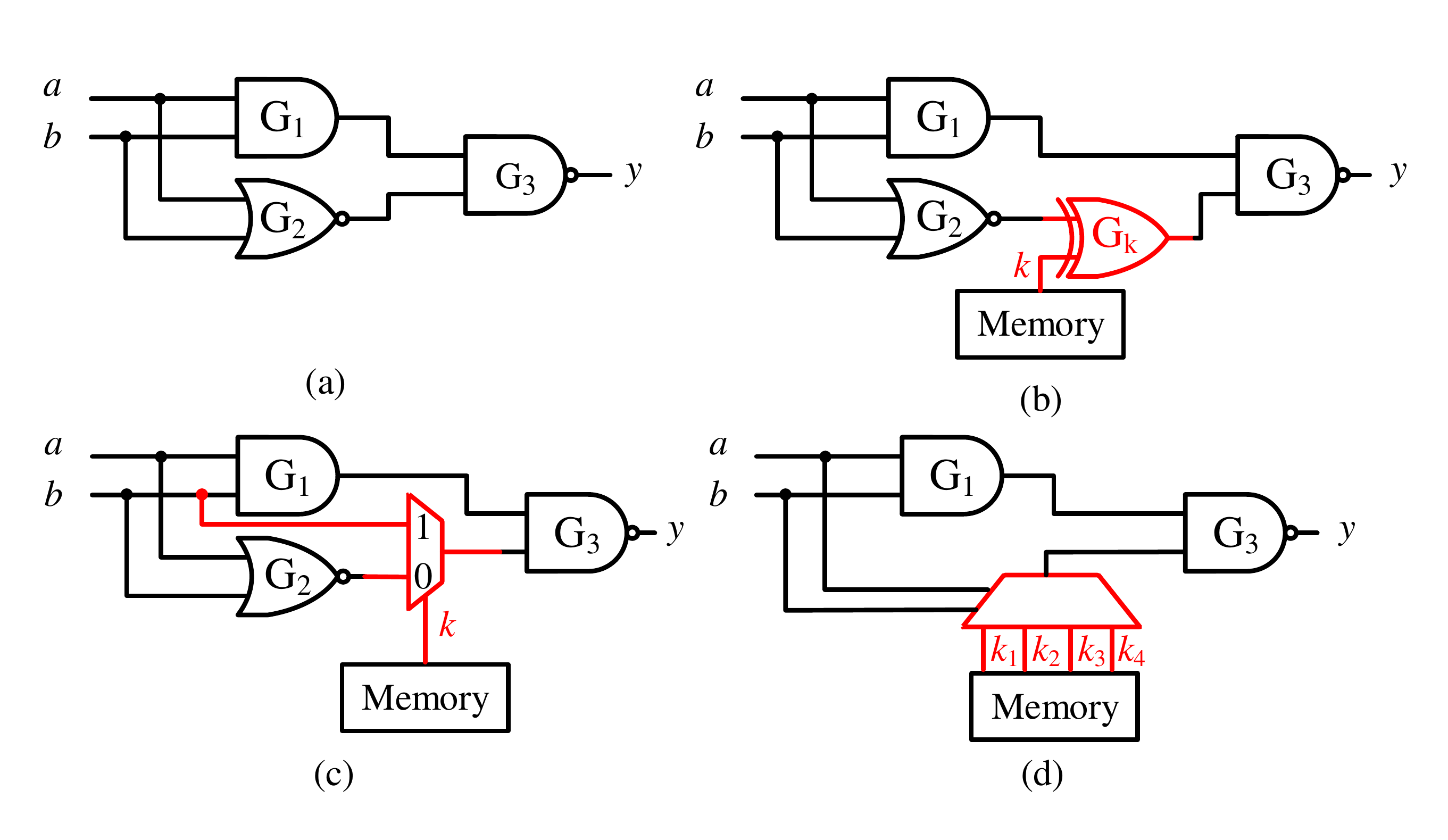} \vspace{-20px}
\caption{ Logic locking methods: (a) An original netlist (b) XOR/XNOR-based logic locking (c) MUX-based logic locking (d) LUT-based logic locking. }  \label{fig:logic-locking}
\vspace{-20px}
\end{figure}

Among the many, logic locking is the most widely accepted and studied design-for-trust technique to prevent threats from untrusted manufacturing and testing. Logic locking hides the circuit's inner details by incorporating key gates in the original circuit resulting in a key-dependent locked counterpart. The resultant locked circuit functions correctly once the secret key is programmed in its tamper-proof memory. Otherwise, it will produce erroneous outputs for the same input patterns, which makes it practically unusable. Over the years, different locking techniques are proposed, which can be primarily categorised based on key-insertion strategy (see Figure~\ref{fig:logic-locking}) and can be described as -- $(i)$ XOR-based~\cite{Roy4484823, RajendranDAC2012, GuinTODAES2016, GuinTVLSI2018, GuinVTS2017}, $(ii)$ MUX-based~\cite{rajendran2015fault, plaza2015solving, lee2015improving}, $(iii)$ LUT-based~\cite{baumgarten2010preventing, khaleghi2015ic, liu2014embedded}, and $(iv)$ state-space based~\cite{chakraborty2009harpoon}. However, XOR-based logic locking is popular due to its simplicity. 

The research community is continuously driven to reveal logic locking vulnerabilities through attacks and to propose countermeasures in turn. The majority of early work was demonstrated vulnerable by oracle guided key-pruning attacks~\cite{SubramanyanHost2015} and its variants~\cite{shen2017double, shamsi2017appsat, alrahis2019scansat2, shamsi2019ip}. Since then, many SAT resilient solutions have been proposed~\cite{wang2017secure, sengupta2020truly, GuinTODAES2016, GuinVTS2017, GuinTVLSI2018, karmakar2018encrypt, potluri2020seql, chiang2019looplock, juretus2019increasing, shakya2020cas}. However, some of them have been broken as well~\cite{sirone2020functional, limaye2019robust, alrahis2019scansat2, shamsi2019kc2, shakya2020defeating, chen2019genunlock, tan2020efficacy}. Even though the SAT attacks are widely popular amongst the research community, the attack model assumes the availability of an oracle or a functionality correct~(unlocked) IC pre-loaded with the correct key, and the adversary has the scan-chain access to obtain the input-output responses. This serves as the limitation as many of the chips used in critical or DoD applications are highly unlikely to be circulated (unless it is a commercial-off-the-shelf, COTS part) in the market right after manufacturing. In addition, the concept of restricting scan-access has also been adopted to provide security against the SAT attacks. An adversary is not restricted to perform only SAT-based attacks as it may deploy other effective attacks to extract the secret key from a locked netlist. Therefore, it is necessary to consider and explore the different directions by which an untrusted foundry can exploit security vulnerabilities to undermine the security of logic locking.

\vspace{-5px}
In this paper, we propose a novel oracle-less attack on logic locked circuits to determine the key. Exploring the capabilities of an adversary, \textit{\textbf{is it possible to determine the secret key simply by analyzing the circuit topology?}} The answer is yes, as the entire circuit topology is built from basic Boolean functions that are repeated multiple times. An adversary can determine the secret key by comparing the locked instances of these functions with the unlocked instances in the entire netlist. This proposed attack is an oracle-less self-referencing attack. We denote our proposed attack as \textbf{\textit{TGA}}: {\textit{T}}opology-\textbf{\textit{G}}uided \textbf{\textit{A}}ttack on logic locked circuits. By using our proposed attack, the secret key can be estimated efficiently even for the circuits that the SAT attack fails (see in Section~\ref{sec:result-and-discussion} for \textit{c6288} circuit). In addition, an adversary can unlock any netlist using our proposed attack without waiting for a working chip available in the market or with no scan access. This was further validated and demonstrated at \textit{UF/FICS Hardware De-obfuscation competition at Trusted and Assured Microelectronics~(TAME) forum~\cite{Competition, TAME}}. The contributions of this paper are as follows:

\begin{enumerate}
    \item \textit{A novel oracle-less topology-guided attack on logic locking:} We proposed a topological function search attack that relies on identifying and searching the repeated functions in a netlist. We denote these basic functions as unit function~\textit{UF}, which are repeated multiple times in a circuit. If a key gate is placed in an instance of repeated \textit{UF} during the locking of a circuit, the original netlist can be recovered by searching the equivalent unit functions (\textit{EUFs}), which are constructed with all hypothesis key values. As the \textit{UFs} are constructed in few layers of gates, the number of key gates and key bits associated with a \textit{UF} is limited, resulting in minimal \textit{EUF} search combinations. The results in Table~\ref{table:prediction3} show the efficiency of the proposed attack by recovering the majority of key bits correctly for ISCAS'85 and ITC'99 benchmark circuits locked with Random Logic Locking~(\textit{RLL}) and Secure Logic Locking~(\textit{SLL}). The effectiveness of our proposed \textit{TGA} is also validated using locked benchmarks from TrustHub~\cite{TrustHub} (see Table~\ref{table:SLL}). In contrast with the traditional oracle~(unlocked chip) attacks, no oracle is required to launch our proposed attack. 
    
    \item \textit{An efficient function search algorithm:} To perform the search, an efficient Depth-First-Search (\textit{DFS}) based algorithm is developed to find the equivalent unit functions in a locked netlist. The complete netlist is first converted to a directed graph~\cite{tarjan1972depth}, where each gate in the netlist is represented as a vertex, and each wire is modeled as an edge. This paper demonstrates and implements a \textit{DFS}-based \textit{EUF} search algorithm to determine the correct value of a secret key. The average time to determine a secret key bit is in the order of seconds. As a result, a locked circuit can be broken in a few minutes, locked with a few hundred/thousand key gates. 
    
    \item \textit{A countermeasure against the proposed \textit{TGA} attack:} As the proposed attack recovers the original design by performing the \textit{EUF} search in the netlist, it can be prevented if the function search with hypothesis keys does not find results or produces contradictory results. This resiliency against the attack can be achieved by inserting the key gates in all the repeated instances of an \textit{UF} as the adversary will not decide the actual value of the key bit by comparing it with its unlocked version. \textit{DFS}-based search algorithm is again exploited to identify all repeated and unique instances of a unit function. Note that the key length can be variable in a range instead of a fixed value, which can increase both the efficiency of the key insertion and the security of the locked design.
\end{enumerate}

The rest of the paper is organized as follows: the background of XOR-based logic locking is provided in Section~\ref{sec:background}. We present our proposed topology-guided attack methodology in Section~\ref{sec:attack-function-search}. We present the countermeasure against the proposed attack in Section~\ref{sec:countermeasure}. We present the results for the implementation of the proposed attack on different logic locked benchmark circuits in section~\ref{sec:result-and-discussion}. Finally, we conclude our paper in Section~\ref{sec:conclusion}. 

\vspace{-10px}
\section {XOR-based logic locking}\label{sec:background}
\vspace{-5px}
To describe our proposed topology-guided attack based on function search, it is necessary to present XOR-based logic locking. Additionally, we need to analyze the resulting circuit modifications based on the selected correct key bit and the key gate type (either XOR or XNOR) to lock the original functionality. This will assist in building equivalent unit functions~(\textit{EUFs}) that will be searched in the netlist to perform the proposed attack.


\begin{figure}[ht]
\centering \vspace{-20px}
\includegraphics[width=1.0\linewidth]{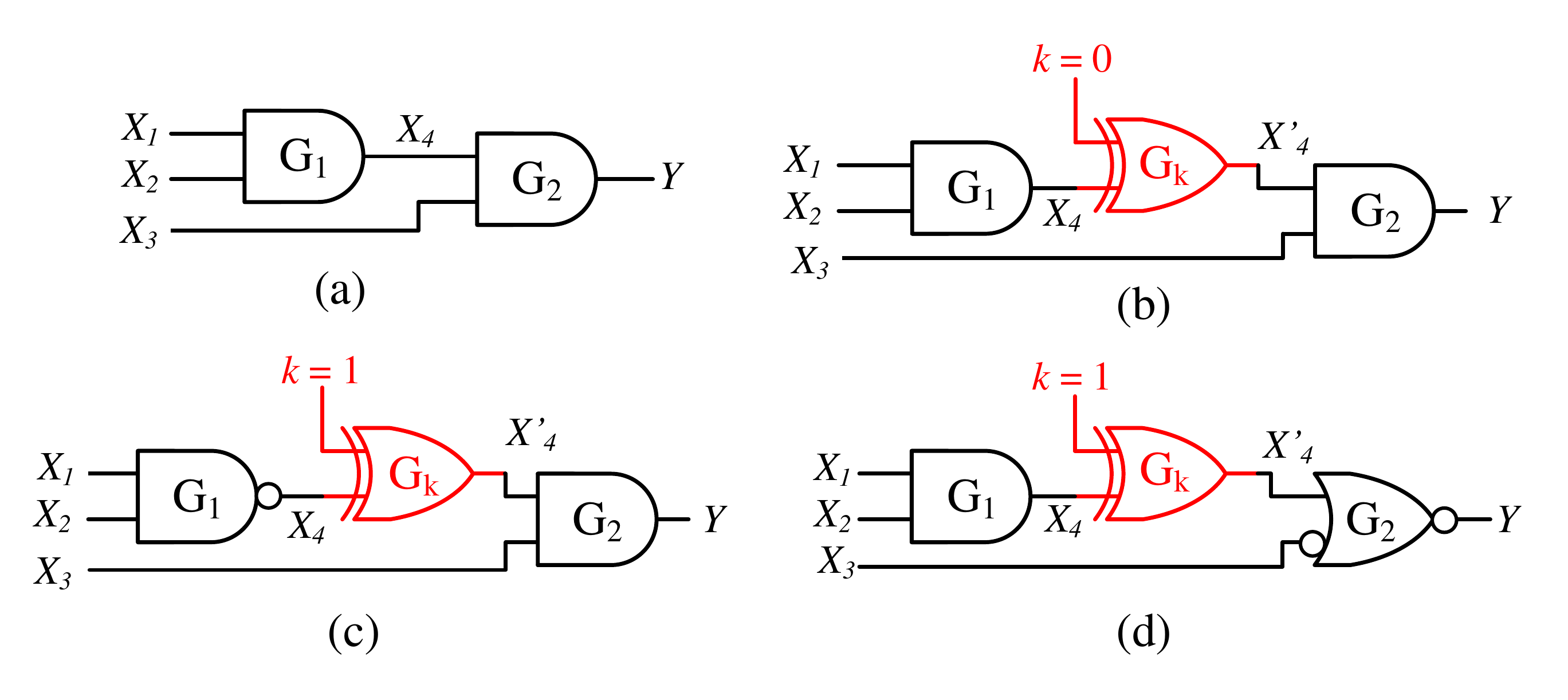} \vspace{-20px}
\caption{Logic locking using Exclusive OR (XOR) gates. (a) Original netlist. (b) Locked netlist when $k$ = 0. (c) \textit{Case-I}: Locked netlist when $k$ = 1. (d) \textit{Case-II}: Locked netlist when $k$ = 1 (using DeMorgan's Theorem).} 
\label{fig:flipfunction}
\end{figure}
\vspace{-15px}

Figure~\ref{fig:flipfunction} shows an example to lock a circuit using an XOR gate, which has three inputs ($X_1$, $X_2$ and $X_3$) and one output ($Y$). One key gate with value $k$ is selected to obfuscate the functionality of the circuit. The original circuit is shown in Figure~\ref{fig:flipfunction}.(a). There can be two possible key values, $k=0$ and $k=1$. For $k=0$, an XOR gate can directly be placed at node $X_4$, which is shown in Figure \ref{fig:flipfunction}.(b). However, for $k=1$, two possible scenarios may occur. One can invert the previous stage functionality, which is shown in Figure \ref{fig:flipfunction}.(c). It is also possible to modify successive stage function using DeMorgan's Theorem, shown in Figure~\ref{fig:flipfunction}.(d).    

In this example, the original function of the circuit is $Y=X_3 \cdot X_4$, where $X_4=X_1\cdot X_2$. It is not necessary to change the functionality of the preceding or succeeding stages of the XOR gate, when $k$ = 0. 

\vspace{-10px}
\begin{equation}
  X^{'}_{4}=X_4 \oplus 0=X_4=X_1\cdot X_2 
\end{equation}
  
To preserve the original functionality for $k$ = 1, it is required either to invert the functionality of the preceding stage (Figure~\ref{fig:flipfunction}.(c)) or compensate the functionality of the following stage (Figure~\ref{fig:flipfunction}.(d)) of the added XOR gate. For the first case, the original functionality preserves as $X^{'}_{4}=1 \oplus \overline{X_4}=X_4$. For the second case, DeMorgan's transformation is necessary as shown below: 

\vspace{-10px}
\begin{equation}
     Y=\overline{\overline{X_3}+X^{'}_{4}}=\overline{\overline{X_3}}\cdot \overline{X^{'}_{4}}=X_3\cdot \overline{(1 \oplus X_4)}=X_3\cdot X_4
\end{equation}

Note that only XOR gates are used in the example to lock the netlist. However, one can also use XNOR gates for such purposes, which has the opposite logic function compared with the XOR gate. It is important to remember that one cannot insert the XOR gate with $k = 0$ and XNOR gate with $k = 1$ for every key bit, as the adversary can determine the secret key just by observing the type of key gates. 

\vspace{-5px}
\section {Proposed Topology-Guided Attack on Logic Locking} \label{sec:attack-function-search}
\vspace{-5px}
The general locking strategy adopted to provide security in a circuit includes the placement of key gates either randomly or in some particular manner~(\eg pair-wise). Since, the secret key associated with the key gates is the same for all the chips manufactured with the same design, finding this key from one netlist undermines the security resulted from logic locking. In this section, we show how an adversary can easily extract the secret key for a key-based locked design using our proposed oracle-less and topology-guided attack, which is built on searching the hypothesis key-based equivalent unit function in the entire locked netlist. Moreover, this attack overcomes the limitations of SAT attacks that require an oracle with the scan access. For the same, we present the different steps involved in performing the proposed attack.

\vspace{-10px}
\subsection{Adversarial Model}\label{subsec:attack-model}
The specific objective is to undermine the security of a logic locking technique by determining the secret key. The secret key is stored in a secure and tamper-proof memory so that the adversary cannot access the key values directly from an unlocked chip. The adversarial model is presented to clearly state the resources and the assets possessed by an adversary. In our attack model, the adversary is assumed to be an untrusted foundry and has access to the following: 

\begin{itemize}
    \item \textit{Gate-level netlist:} As the primary attacker, the foundry can have access to the gate-level netlist of a locked IC. The SoC designers typically send the circuit layout information using GDSII or OASIS files~\cite{reich2003oasis} to a foundry for chip fabrication. With the help of advanced tools, the foundry can extract the gate-level netlist from those provided GDSII/OASIS files~\cite{torrance2009state}.
    \item \textit{Location of the key gates:}  The location of key gates can be determined as these gates are connected either directly or through temporary storage elements to the tamper-proof memory. An adversary can easily track the routing path from the tamper-proof memory to the corresponding gates to determine their locations. 
    \item \textit{Locked unit function:} It is trivial for an untrusted foundry to construct equivalent unit functions \textit{EUFs} for launching the topology-guided attack, as it has the netlist and locations of the key gates.   
\end{itemize}

\vspace{-15px}
\subsection{Motivation}\label{subsec:Motivation}
\vspace{-2px}

\begin{figure}[t]
\centering
\includegraphics[width=1.0\linewidth]{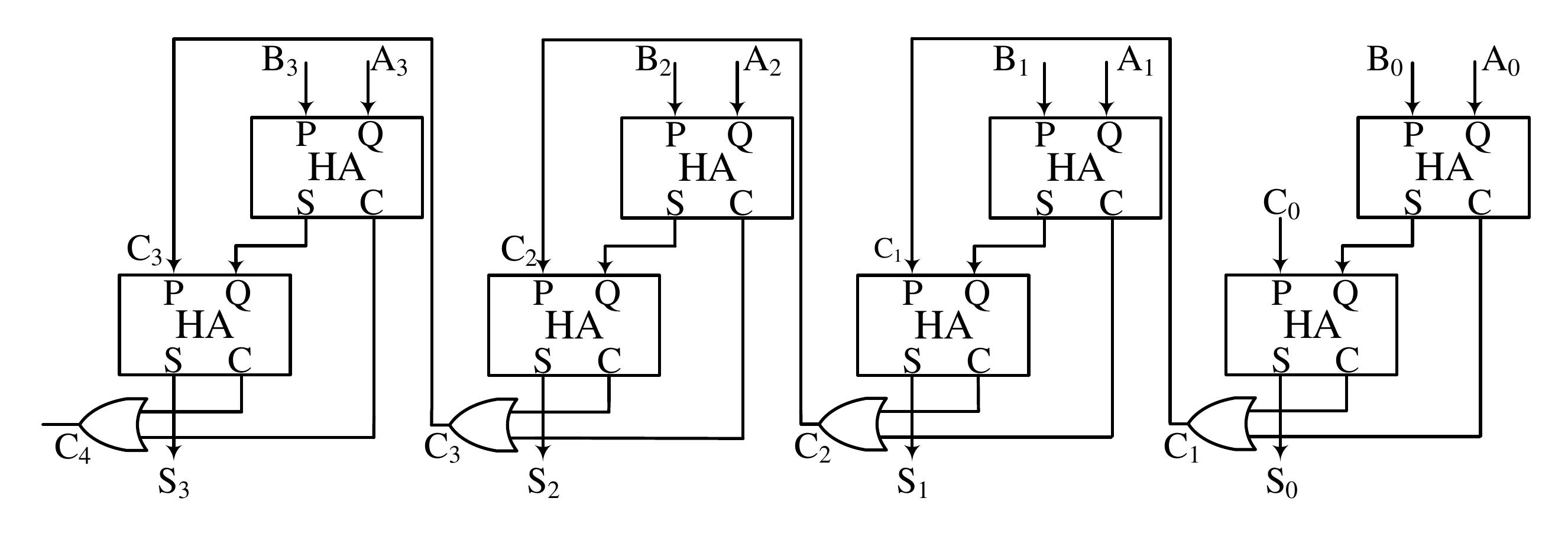} \vspace{-20px}
\caption{Four-bit ripple carry adder consists of eight identical half adders (\textit{HA}). If a \textit{HA} is locked, an adversary can recover the original netlist  by simply comparing it with other unlocked \textit{HAs}.} \label{fig:fulladder}
\end{figure}

The basic idea of launching our proposed attack is based on the repeated functionality that exists in a circuit. The Boolean functions are generally not unique in a circuit and repeated multiple times to implement their overall functionality. The majority of circuits are constructed based on small functional units. For example, several small functions (we describe as `unit functions' or \textit{UFs}) are repeated in an arithmetic logic unit (ALU) of a processor, adders, multipliers, advance encryption standards (AES), RSA, and many other digital circuits. If any of such \textit{UFs} are not obfuscated during logic locking, all the locked functions will be unlocked simply by comparing them with their unlocked version. 

Figure \ref{fig:fulladder} provides a four-bit ripple carry adder circuit as an example to illustrate the concept of our proposed attack. This full adder~\textit{FA} consists of eight identical one-bit half adders (\textit{HA}) with inputs ($P$ and $Q$) and outputs ($S$ and $C$). Each individual half adder can be considered as a unit function \textit{UF}, which is repeated multiple times inside this full adder. If one of these half adders is locked using an XOR/XNOR gate, an adversary only needs to find an original unlocked \textit{HA}, and then match this with the locked HA to recover the key value (see details in Section \ref{subsec:proposed-attack}).

\vspace{-10px}
\subsection{Construction of Equivalent Unit Function} \label{subsec:construction-EUF}
\vspace{-5px}
Our proposed attack constructs an equivalent unit function to perform the search. While constructing the \textit{EUF}, an adversary may encounter two different cases, either there is only one key gate, or there are multiple key gates in the \textit{UF}. In either case, the (\textit{EUF}) is constructed using one/more hypothesis key bits or a combination of hypothesis key bits, and searches that \textit{EUF} in the entire netlist to find a match. The hypothesis key bits will be the correct secret key bits for the respective \textit{UF} if a match is found corresponding to the \textit{EUF}. Otherwise, it constructs another \textit{EUF} using a different combination of values for the hypothesis key bits in both the cases and searches the netlist again. The number of \textit{EUFs} depends on the number of key gates included in the \textit{UF}. In this section, we show how \textit{EUFs} are created to determine the secret key for both \textit{RLL} and \textit{SLL} circuits.

\subsubsection{Random logic locking}
\vspace{-5px}
In random logic locking (\textit{RLL}), the key gates are inserted randomly inside the circuit that needs to be protected. In the large design with thousands of gates, it is highly unlikely that multiple key gates will be inserted adjacent to each other. Thus, the inserted key gates usually can be considered individually to construct the equivalent unit functions.

\begin{figure}[!ht]
\centering
\vspace{-20px}
\includegraphics[width=1.0\linewidth]{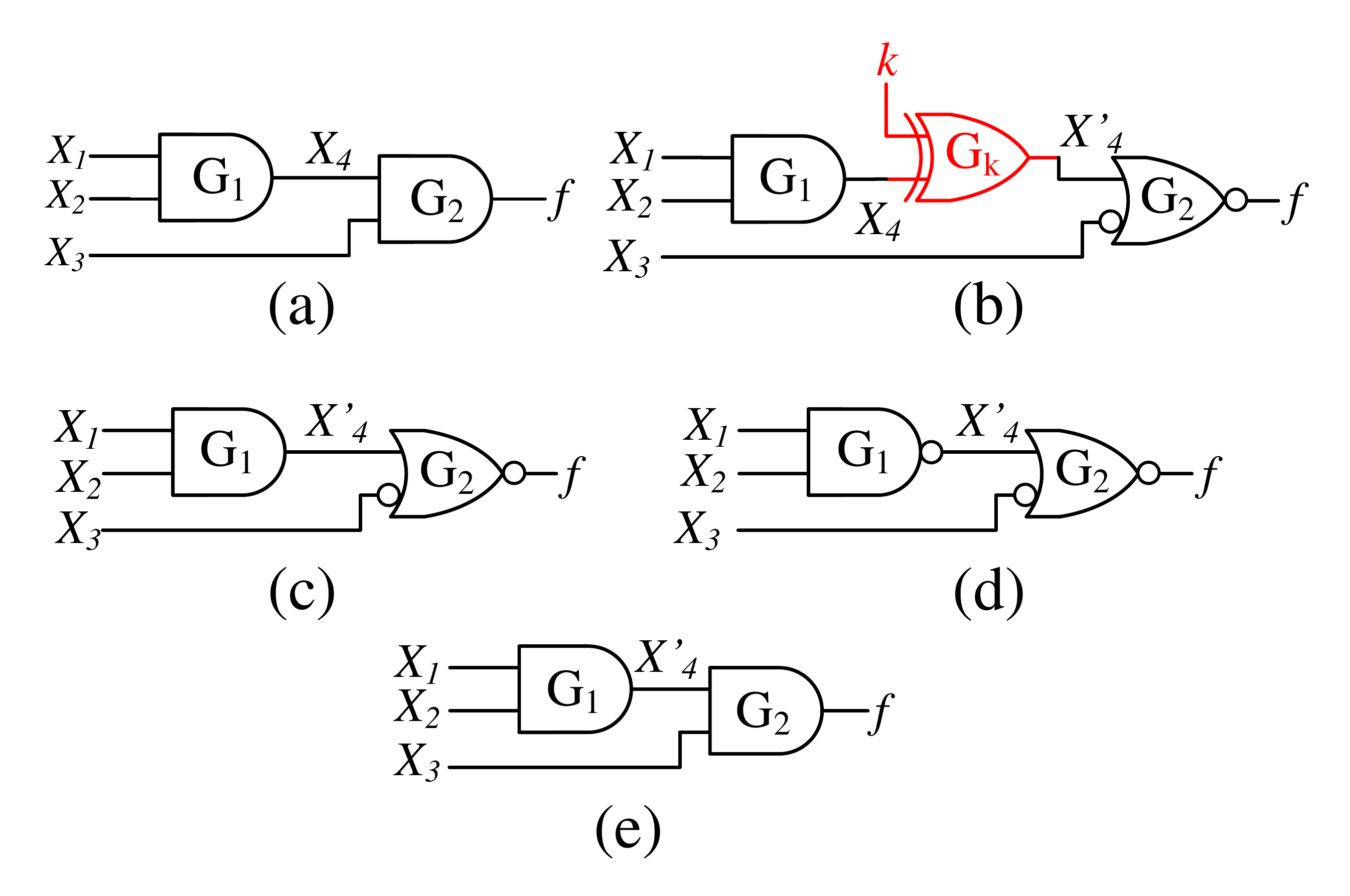} \vspace{-20px}
\caption{\textit{EUF} construction  for different hypothesis key values. (a) Original unlocked netlist. (b) Netlist is secured with key value $k=1$. (c) $EUF_0$ for hypothesis key $k_h=0$. (d) $EUF_1$ for hypothesis key $k_h=1$ (\textit{Case-I}). (e) $EUF_{\hat{1}}$ for hypothesis key $k_h=1$ (\textit{Case-II}).} \label{fig:Equivalent-UF-Hypothesis-Key}
\end{figure}

\vspace{-5px}
Figure \ref{fig:Equivalent-UF-Hypothesis-Key} illustrates the construction of equivalent unit functions with a single key gate, which can be used to launch the function search attack. Figure~\ref{fig:Equivalent-UF-Hypothesis-Key}.(a) represents an original unit function to be locked using a correct secret key $k=1$. The locked circuit is shown in Figure~\ref{fig:Equivalent-UF-Hypothesis-Key}.(b). The adversary cannot deduce the value of the key, simply by observing the key gate. It first makes an assumption for $k_h=0$, and constructs the \textit{EUF}, which is shown in Figure \ref{fig:Equivalent-UF-Hypothesis-Key}.(c). It then searches this function in the locked circuit to find a match. If no match is found (as the actual key is 1), it constructs another \textit{EUF} for $k_h=1$. Two possible scenarios may occur. For \textit{Case-I}, the output of the previous stage needs to be inverted (shown in Figure~\ref{fig:Equivalent-UF-Hypothesis-Key}.(d)). On the other hand, DeMorgan's transformation needs to be carried out to obtain the \textit{EUF} for $k_h=1$ for \textit{Case-II}, which is shown in Figure~\ref{fig:Equivalent-UF-Hypothesis-Key}.(e). As inferred from the construction of the equivalent unit function, each key gate has two hypothesis keys with three transformations represented as: (i) $EUF_0$ where the hypothesis key $k_h=0$, (ii) $EUF_1$ (\textit{Case-I}) where the hypothesis key $k_h=1$, and (iii) $EUF_{\hat{1}}$ (\textit{Case-II}) where the hypothesis key $k_h=1$ but the modification is carried out using DeMorgan's Theorem. Three \textit{EUFs} can be constructed for a single key bit. For a hypothesis key bit $k_h = 0$, a single implementation of \textit{EUF} can be considered, whereas, two different implementations can be possible for hypothesis key bit $k_h = 1$. As a result, for an \textit{UF} locked with a $j$-bit key, $3^j$ number of implementations can be possible. We need to perform all $3^j$ \textit{EUF} search to determine the $j$-bit key.




\vspace{-5px}
\subsubsection{Strong logic locking}
\vspace{-5px}
The objective of strong logic locking (\textit{SLL}) is to maximize the interference between different key gates to restrict key sensitization at the output~\cite{yasin2016improving}. In \textit{SLL}, two or more key gates are inserted adjacent to each other so that their outputs converge at the next stage logic gate. The propagation of one of the key-bit will be possible only if certain conditions are forced on other key inputs or they are known. As these key inputs are not accessible by the attackers, they cannot force the logic values necessary to sensitize a key. As a result, the proposed \textit{TGA} on \textit{SLL} requires an equivalent unit function search with multiple keys instead of a single one for random logic locking.

\begin{figure}[!ht]
\centering
\vspace{-10px}
\includegraphics[width=1.0\linewidth]{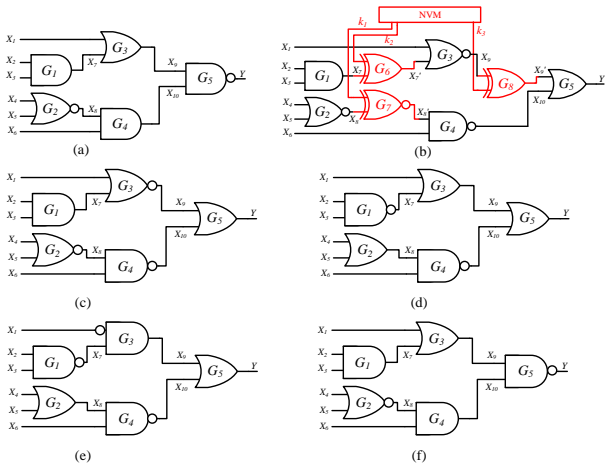} \vspace{-20px}
\caption{Equivalent unit functions for multiple gates with  different hypothesis keys. (a) Original netlist. (b) Locked netlist with key value $k_1k_2k_3=101$. (c) $EUF_{100}$ for hypothesis key, $k_h=100$. (d) $EUF_{011}$ for $k_h=011$. (e) $EUF_{0\hat{1}0}$ for $k_h=010$. (f) $EUF_{10\hat{1}}$ for $k_h=101$.} 
\label{fig:multiple_gates}
\vspace{-10px}
\end{figure}

Figure~\ref{fig:multiple_gates} illustrates the construction of \textit{EUFs} with multiple key gates that will assist in performing the function search. The original unit function (as shown in Figures~\ref{fig:multiple_gates}.(a)) is locked with three key gates to increase the inter key-dependency. The locked unit function is shown in Figure~\ref{fig:multiple_gates}.(b) with correct key $k_1k_2k_3=101$. As an adversary cannot extract the correct key value from the non-volatile memory directly, all the \textit{EUFs} will be constructed and searched in the entire locked netlist. However, the number of constructed \textit{EUFs} will increase due to the number of key gates and its combination in the locked \textit{UF}. As mentioned earlier, each key gate results 3 different \textit{EUFs}~(e.g., $EUF_0$, $EUF_1$, and $EUF_{\hat{1}}$ based on the hypothesis key values (either 0 or 1). This will result in overall 27 \textit{EUFs} (\ie $3^3$, as $j=3$ is the number of key gates in the \textit{UF}) for Figure~\ref{fig:multiple_gates}.(b), amongst which only 4 of them are shown in Figure~\ref{fig:multiple_gates}.(c)-(f). These \textit{EUFs} are derived from different key combinations. For example, $EUF_{100}$ in Figure~\ref{fig:multiple_gates}.(c) is constructed with hypothesis key bits based transformation as $EUF_1$, $EUF_0$ and $EUF_0$ for key gates $G_7$, $G_6$ and $G_8$ respectively. Also, we construct $EUF_{011}$ as shown in Figure~\ref{fig:multiple_gates}.(d). if we transform based on the hypothesis key bits $k_h=011$ for key gates $G_7$~($EUF_0$), $G_6$~($EUF_1$) and $G_8$~($EUF_1$). Figure~\ref{fig:multiple_gates}.(e) shows yet another \textit{EUF} represented as $EUF_{0\hat{1}0}$, where the hypothesis key is 0, 1~(Case-II) and 0.  Likewise, if we select the transformation as $EUF_1$, $EUF_0$ and $EUF_{\hat{1}}$~(Case-II) for key gates $G_7$, $G_6$ and $G_8$ respectively, then we will get the $EUF_{10\hat{1}}$ shown in Figure~\ref{fig:multiple_gates}.(f). Once all the \textit{EUFs} are constructed, all of them will be searched in the netlist to find a match. As Figure \ref{fig:multiple_gates}.(f) is identical to Figure \ref{fig:multiple_gates}.(a), the hypothesis key combination $k_h=101$ should be the correct key value. If no such match is found for any of the \textit{EUFs}, an adversary cannot make the prediction on the key combination resulting the \textit{UF} being unique in the circuit.

\vspace{-10px}
\subsection{Function Search using \textit{DFS} Algorithm} \label{subsec:SUF}
\vspace{-5px}
An efficient search algorithm has been developed to search the \textit{EUFs} in the locked netlist. The structure of a circuit can be transformed and represented as a directed graph, and all the algorithms that can be used to search the component in the directed graphs, can also be applied to search the \textit{EUF}. Therefore, we propose to use the Depth-First-Search (\textit{DFS})-based algorithm to launch the attack. Generally, the \textit{DFS} method follows the rule: in the graph traverse procedure, the edge from the most recently reached and connected vertex that still has unexplored edges will always be selected as next edge~\cite{tarjan1972depth}. Before performing the \textit{DFS}-based search, a data object structure needs to be defined to store and transform the netlist as a directed graph. The gate object needs to have the following attributes: gate type (\eg XOR, AND, etc.), name of the gate (i.e., its identification in the netlist), an array that contains its preceding gates (\ie its inputs), and an array contains its following gates (\ie its outputs). Then the circuit structure can be transformed and stored into a dictionary, in which the keys are the types of the gates and the values are corresponding gate objects. Dictionary is basically a data structure that stores mappings and relationships of data~\cite{cormen2009introduction}. The use of a dictionary makes the search for specific type of gates more efficient.

\setlength{\textfloatsep}{10pt}
\begin{algorithm}[h]
  \caption{\textbf{Function \textit{FS}} \protect\\
  Function search based on \textit{DFS} Algorithm.} \label{alg:dfs}
  \SetKwInOut{Input}{Input}\SetKwInOut{Output}{Output}
  \SetKw{Break}{break}
  \SetKwFunction{FMain}{\textit{DFS}}
  \SetKwProg{Fn}{Function}{:}{\KwRet $F$}
  \Input{The gate-level netlist of a circuit ($C$), Function ($Fn$)}
  \Output{Result List ($L_R$) } 
  
  \algrule
  
  Read $C$ and $Fn$, and transform them into dictionaries, $O$ and $T$\; 
  
  $R \leftarrow  Fn.root$; $L_S \leftarrow  O[R.type]$; $L_R \leftarrow  \phi$ \; 
  
  \For{each gate $G$ in $L_S$}{
  \If{$DFS(R,G)$}{$L_R$.\textit{append}($G$)\;}
  }
  
  {\KwRet} $L_R$\; 
  \BlankLine
  \Fn{\FMain{$r,g$}}{
  
  $F \leftarrow$ True\;
  $L_1 \leftarrow r$.\textit{PrecedingGates};  $L_2 \leftarrow g$.\textit{PrecedingGates}\;
  $T_1 \leftarrow L_1$.\textit{types};  $T_2 \leftarrow L_1$.\textit{types}\;
  \If {$L_1$ is empty}{\Return \textit{True}\;}
  
  \For{each gate type $T$ in $T_1$}{
  \uIf {gate type $T$ not in $T_2$} {\Return \textit{False}\;}
  \uElse{$T_2$.remove($T$)}
  }

  \For {each gate $R_N$ in $L_1$}{
  
   $L_T \leftarrow \phi$ \;
  
  \For{each gate $G_T$ in $L_2$}{
  
  \If{$G_T$.type = $R_N$.type}{$L_T$.\textit{append}($G_T$)\;}
  }
  $F_T \leftarrow$ \textit{False}\;
   
  \For{each gate $G_N$ in $L_T$}{
 
  \If{DFS($R_N, G_N$)}{
  $F_T \leftarrow$ \textit{True}\;
  
  \Break}
  
  }
  $F \leftarrow F * F_T$\;
  }
  } 
\end{algorithm} 


The procedure of \textit{DFS}-based search is described in Algorithm~\ref{alg:dfs}. The function \textit{FS} finds matches for any function (\textit{Fn}) in a circuit that it takes as an input. Whenever a specific \textit{Fn} need to be searched in this netlist, we define the last gate of the \textit{Fn} as the root gate (Line 2 in the Algorithm~\ref{alg:dfs}). An example root gate is $G_2$ in the Figure~\ref{fig:Equivalent-UF-Hypothesis-Key}). All the gates that have the same type with the root gate ($G_2$) in the dictionary (Line 3) are stored into an array. The \textit{DFS} is then performed on all these found gates (Line 3-7). Finally, all the \textit{UFs} in the netlist will be found and the count of the \textit{F} will be returned as the output (Line 8). The detailed implementation of the \textit{DFS} is demonstrated in Lines 9-38.

The algorithm is implemented with Python 2.7~\cite{Python-2.7}. The worst case time complexity of the search algorithm is $O(n*u)$, where $n$ is the size of the netlist and $u$ is the size of a unit function. This is an acceptable complexity, since it is known that the subgraph isomorphism problem is an NP-complete problem and its time complexity is quadratic in the number of nodes~\cite{reich2003oasis, dickinson2003graphs}. Note that the optimization of the algorithm complexity is not the major objective of this paper. However, our search strategy slightly reduces the search complexity by using a dictionary to locate root gates. In this case, the algorithm performs similar to a subtree isomorphism search (or a sequence of tree isomorphism searches), whose complexity is known to be at least subquadratic~\cite{abboud2018subtree}. Reading the netlist and transforming it into a dictionary may have different complexity. The complexity analysis does not consider the complexity of constructing a netlist dictionary. 

\subsection{Proposed attack using Equivalent Unit Function Search} \label{subsec:proposed-attack}
\vspace{-5px}

\setlength{\textfloatsep}{10pt}
\begin{algorithm}[!b]
  \caption{Topology-guided attack using \textit{FS}} \label{alg:prediction}
  \SetKwData{Left}{left}\SetKwData{This}{this}\SetKwData{Up}{up}
  \SetKwFunction{Union}{Union}\SetKwFunction{FindCompress}{FindCompress}
  \SetKwInOut{Input}{Input}\SetKwInOut{Output}{Output}
  \SetKwFunction{FMain}{FV}
  \SetKwProg{Fn}{Function}{:}{\KwRet $k_i$}
  \Input{Locked Circuit Netlist $(C^*)$}
  \Output{List of predicted key values $(K_P)$, Success Rate ($SR$)}
  
  \algrule
  
  Read the netlist $C^*$\;
  
  Determine the location and number $|K|$ of key gates\;
  
  Initialize correct prediction counter, $p_c \leftarrow 0$ \;
  
  \For{$i\leftarrow 1$ \KwTo $|K|$}{
  
  \uIf{$k_i$ is not determined in $K_P$}{ 
  
  Initialize layer counter, $l \leftarrow 1 $\;
  
  Get the unit function for $k_i$ based on $l$ \;
  
  Get number of key gates $j$\ in the function\;
  
  $J \leftarrow$ key combinations list, where the length of $J$ = $2^j$\;
  
  $R \leftarrow [0]*2^j$ \; 
  
  Generate $3^j$ equivalent unit functions for the $j$-bit key   \;
  
  \For{each generated EUF}{
  
  $J'\leftarrow$ hypothesis key of EUF\; 
  
  $R[J.index(J')] \leftarrow R[J.index(J')] + FS(C^*, EUF).sz()$\;
  
  }
  
  \uIf{$R.nonzero$ = 1}{ 
  $J' \leftarrow R.index(1)$\; 
  Correct hypothesis key $k_j \leftarrow J[J']$ \;
  \If{Any key gate in $k_j$ is placed in a fan-out net}{
  $k_j=FV(~)$\;
  }
  Write $k_j$ into $K_P$; $p_c \leftarrow p_c + j$\;
  }
  \uElseIf{$R.nonzero$ = 0}{
  $k_1...k_j \leftarrow $ $X$\;
  Write $k_1...k_j$ into $K_P$;
  }
  \uElse{$l \leftarrow l + 1$, go to line 7} 
  
  }
  \uElse{Continue}
  }
  
  Compute success rate, $SR \leftarrow \frac{p_c}{|K|} \times 100\%$\;
  Output $K_P$, $SR$\;  
  \BlankLine
  \Fn{\FMain{$~$}}{
  Construct different \textit{EUFs} for the fanout paths\;
  Search \textit{EUFs} for each path and make key prediction \;
  \uIf{Opposite predictions for different paths}{ 
  $k_i\leftarrow$X \;
  }  
  \ElseIf{Same predictions for different paths}{
  $k_i\leftarrow \{0 \text{ or } 1\}$\;
  }
  } 
\end{algorithm} 

The objective of the proposed topology-guided attack is to recover the entire original netlist using the equivalent unit function search (\textit{FS}). Algorithm~\ref{alg:prediction} describes the proposed attack. The locked circuit~($C^*$) is given as the input, and the list of predicted key values~($K_P$) with the success rate~(\textit{SR}) will be returned as outputs. $K_P$ contains the predicted value of each key gates, which can be either 0, 1, or X. The X represents an unknown value when the search fails to find a match and make the prediction. The locations of the key gates can be found by tracking the routes originated from the tamper-proof memory, and their numbers can be determined as $|K|$. In order to determine the key value inside a particular unit function, different unit functions need to be constructed based on the number of key gates inserted in this unit function. In addition, each of the key gate comes with a hypothesis key value (either 0 or 1), and this also leads to the different hypothesis key combinations when there are multiple key gates inserted in a \textit{UF}. 

For each key gate $k_i$, the unit function will be constructed based on the value $l$. Here, $l$ denotes how many layers of gates are considered when constructing the unit functions. The $l$ is initialized as 1 at the beginning (Line 6), which is also shown in Figure~\ref{fig:Equivalent-UF-Hypothesis-Key}. Next, the unit function based on the $k_i$ and $l$ will be generated (Line 7), and the number of key gates (includes $k_i$) in this unit function will be determined as $j$ (Line 8). The hypothesis key combinations for all the key gates in this unit function will be generated and stored in a list $J$ (Line 9). Note that the order of the keys has no relationship with the real sequence in the circuit, and the number of the combinations is $2^j$. Once the key combination list is generated, all the possible \textit{EUFs} will be constructed based on the hypothesis key combinations (Line 11). For each key gate, three different cases need to be considered (see Figure~\ref{fig:Equivalent-UF-Hypothesis-Key} for details), thus $3^j$ \textit{EUFs} will be generated. The function search (\textit{FS}) (described in section \ref{subsec:SUF}) is then performed to find the repeated instances of \textit{EUFs} (Line 12-14). $2^j$ count values will be accumulated in a list $R$ (initialized with all 0 in Line 10) for all key assumptions. 

Upon finishing the search of all the \textit{EUFs}, if only one count value in $R$ is non-zero, this non-zero value corresponding \textit{EUF} represents a correct key prediction. The hypothesis key $J'$ of this \textit{EUF} will be written into $K_P$, and the prediction counter ($p_c$) will be increased by the length of this hypothesis key, $j$ (Line 16-22). Note that, if the key gate is placed in a fan-out net, an additional process needs to be performed (Line 19-21). Function $FV(~)$ verifies the key decision on each path. It may happen that different paths for the same key gate may have different key predictions. As a result, no prediction will be made in case of any two (or more) paths provides the opposite key value predictions (Line 36-37). Correct predictions will only be made if different paths make the same prediction (Line 38-39). 

On the other hand, if all of the elements in $R$ are equal to $0$, this means this unit function is unique in the circuit and the adversary cannot make a prediction on the key value. As a result, unknown value (X) is assigned to all the $j$ key gates in this unit function, and the values are also stored in to $K_P$ (Line23-25). In the case of multiple count values in $R$ are non-zero, the adversary can neither make the key value prediction based on the current \textit{EUF}. It is necessary to increase the size of the \textit{EUF} by increasing the layer of gates considered in \textit{EUF} constructions. Therefore, the $l$ value needs to be increased by 1, and the entire searching procedure will be re-performed (26-28).


\vspace{-10px}
\begin{equation} \label{eqn:SR}
    SR =  \frac{p_c}{|K|} \times 100\%
\end{equation}

Finally, the success rate is computed using Equation~\ref{eqn:SR}. Here, $|K|$ presents the size of the key while $p_c$ indicates the value stored in the correct prediction counter.  The algorithm will finally report predicted key list $K_P$ and \textit{SR} (Line 27).

The proposed attack may also cause incorrect predictions. For example, it is possible that the actual key bit is 1 when the attack gives an estimation as 0, and vice versa. It is thus necessary to measure the accuracy of the proposed attack. The misprediction rate (\textit{MR}) of our proposed attack can be described as the ratio of the incorrect predictions to the key size and is presented using the following equation:
\vspace{-5px}
\begin{equation} \label{eqn:misprediction}
    MR =  \frac{p_i}{|K|} \times 100\%
\end{equation}
where, $p_i$ denotes the total number of incorrect predictions.

\vspace{-5px}
\section{Countermeasure for \textit{TGA}} \label{sec:countermeasure} 

In this section, we propose an effective key insertion algorithm, which can prevent the proposed topology-guided attack. As an adversary performs \textit{EUF} search in the netlist to find out the reference \textit{UF}, this attack can be prevented if the search of those key gates and \textit{EUFs} always returns no results or contradictory values. The basic idea of the countermeasure is to lock all the repeated instances of \textit{UFs} and insert the key gate(s) in all unique \textit{UFs} in the circuit simultaneously. As a result, the adversary cannot predict and recover the correct key values by comparing the locked \textit{UFs} with the unlocked version. In order to find all the repeated instances of selected \textit{UF}, the \textit{UF} search will be performed at the beginning before the key gates are placed into the netlist.

\begin{algorithm}[ht]
  \caption{Insertion of key gates to prevent topology-guided attack} \label{alg:TGA-resistant-locking}
  \SetKwData{Left}{left}\SetKwData{This}{this}\SetKwData{Up}{up}
  \SetKwFunction{Union}{Union}\SetKwFunction{FindCompress}{FindCompress}
  \SetKwInOut{Input}{Input}\SetKwInOut{Output}{Output}
  \Input{Gate level netlist of a circuit $(C)$, \\ Key size ($\langle K_{min}, K_{max}\rangle$)}
  \Output{Locked netlist $(C^*)$ and Key value $(K^*)$ }
  
  \algrule
  Initialization: $n \leftarrow 0$, $r \leftarrow 0$\;
  \While{$n < K_{min}$}{
    Select a root gate randomly from $C$\;
    Construct the unit function, $UF$\;
    $r \leftarrow FS(C, UF).sz()$\;
    \uIf{RLL}{
    \uIf{$r = 1$}{
     Insert the key gate at one random input of root gate and assign key value, $k_n \in \{0,1\}$\; 
     Write key value, $K^*[n] \leftarrow k_n$\;
     $n\leftarrow n+1$\;
     }
    \ElseIf{$1 <r\leq K_{max}-n$}{
    Lock all the \textit{UFs}\;
    Write key values to $K^*[n+r:n]$\; 
    $n\leftarrow n+r$\;
    }
    }
     \ElseIf{SLL}{ 
     \uIf{$r = 1$}{
     Insert $j$ key bits in the unique $UF$, and assign key values, $k_n,~k_{(n+1)},~\ldots,~k_{(n+j)}$ \; 
     Write key value, $K^*[n+j:n] \leftarrow [k_{(n+j)},\ldots,~k_{(n+1)},k_{n}]$ \;
     $n\leftarrow n+j$\;
     }
     \ElseIf{$1 <r\leq K_{max}-n$}{
    Lock all the \textit{UFs}\;
    Write key values to $K^*[(n+r*j):n]$\; 
    $n\leftarrow n+r*j$\;
    }
    }
     
  }
  Output $C^*$ and $K^*$\; 
\end{algorithm} 

Algorithm~\ref{alg:TGA-resistant-locking} illustrates our proposed solution for key gate(s) insertion. The original unlocked netlist ($C$) will be provided as the initial input, along with the key size~($\langle$ $K_{min}, K_{max}$ $\rangle$), which indicates the range of number of key gates that needs to be inserted in the circuit. Finally, the locked circuit netlist ($C^*$) and the secret key $K^*$ will be the outputs of the algorithm. 
Here, $n$ denotes the key index, which is the number of key gates that has been already inserted in the circuit and initialized to be 0 (Line 1). The entire process can be described as follows:  First, a gate is selected randomly from the original unlocked netlist as the root gate (Line 3). Then, the unit function based on the root gate will be created (see Figure~\ref{fig:Equivalent-UF-Hypothesis-Key}.(a)) for the \textit{UF} search (Lines 4). Next, $FS(C,UF).sz()$ returns $r$ , which denotes the number of this selected \textit{UF} repeated in the circuit (Line 5). Depending on the value of $r$, whether 1 (unique) or greater than 1 (repeated), key gate(s) can be inserted in this \textit{UF} in accordance with \textit{RLL} or \textit{SLL} techniques.

For \textit{RLL}, $r=1$ signifies the constructed \textit{UF} is unique, and the \textit{FS} function found only one instance (itself) in the netlist. As a result, a random key gate (either XOR or XNOR) will be inserted before the root gate and the \textit{UF} will be modified randomly based on the key value. After the key gate insertion, the key bit value is written in the respective location of $K^*$, and the value of $n$ will be increased by 1 (Lines 9-10). In the case of $r>K_{max}-n$ which represents that the number of this repeated \textit{UF} is more than the maximum remaining number of key gates we expect to insert, the algorithm will randomly choose a different gate as the new root gate (Line 3). Otherwise, the algorithm will lock all the repeated instances of this constructed \textit{UF} in the circuit (Line 12). The respective key bit locations in $K^*$ are written with the key values (Line 13). Note that it is ineffective to lock all these instances with only one key value, \ie all 0s or all 1s, as the attacker can recover the entire netlist by simply analyzing the type of the key gate. A combination of 1s and 0s (shown in Figures~\ref{fig:flipfunction}) will be a better option in order to provide enough security for the circuit. However, it is mandatory to lock all the repeated \textit{UFs}. Finally, the value of $n$ is increased by $r$.

\begin{figure*}[t]
\centering
\includegraphics[width=1\linewidth]{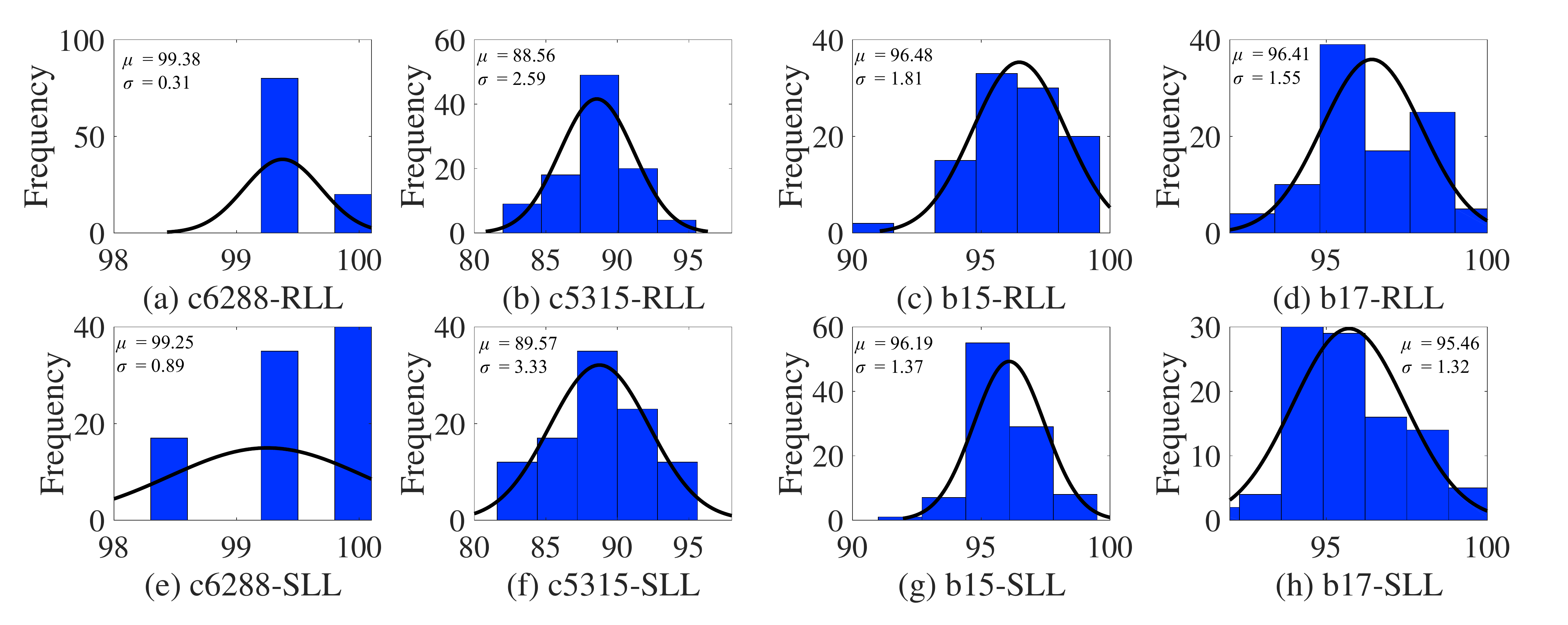} \vspace{-20px}
\caption{Histogram plots of the \textit{SR} for different benchmark circuits with 128 key bits: (a) \textit{c6288-RLL} (b) \textit{c5315-RLL} (c) \textit{b15-RLL} (d) \textit{b17-RLL} (e) \textit{c6288-SLL} (f) \textit{c5315-SLL} (g) \textit{b15-SLL} (h) \textit{b17-SLL}. } \label{fig:SuccessRate}
\vspace{-10px}
\end{figure*}

Similarly, for $r=1$, \textit{SLL} can be carried out by inserting $j$ key gates in the \textit{UF}, namely $k_n, k_{n+1}, \ldots, k_{n+j}$ (line 18). After the insertion of the key gates, the value of these key bits is written in the respective location of $K^*$, and the value of $n$ will be increased by $j$ since $j$ key gates has been inserted already (Lines 19-20). In the case of $r>K_{max}-n$ when the number of this repeated \textit{UF} is more than the maximum remaining number of key gates, the algorithm will automatically choose a different gate as the new root gate (Line 3). Otherwise, the algorithm will lock all the repeated instances of this constructed\textit{UF} with \textit{SLL} (Line 23). The respective key bit locations and values is also updated in $K^*$ (Line 24). At last, the value of $n$ is increased by $r*j$. 

When multiple identical \textit{UFs} are selected, a designer may select to lock these \textit{UFs} with a key of 0 or 1. However, an adversary can identify different locked versions of the same unit functions (\textit{UFs}) and compare them to find the value of the secret key. For example, an adversary can construct the complete truth table of two locked \textit{UFs} with $k_i=0$ and $k_i=1$, and compare their responses. As the correct key bit produces the same responses for both the functions, the key bits can be determined. As a result, if the same repeated \textit{UF} is locked with different key bits, an adversary can determine their value by merely comparing it with the other locked version. An adversary, however, needs to identify the same repeated \textit{UFs} for comparison. As a result, we propose to lock all the repeated UFs with the same key bit so that self-referencing becomes infeasible.
 
Another possible countermeasure against \textit{TGA} can be developed by performing DeMorgan's transformation further from the key gate. When a key gate is inserted in a specific location, the required circuit modification due to the inserted key can be performed further away from it instead of modifying it is preceding or the following logic. As a result, the attacker may fail to make the correct key prediction when the \textit{EUFs} constructed with few gates are searched in the entire circuit. However, an adversary constructs additional possible \textit{EUFs} with DeMorgan's transformation further away from the key gate and searching the netlist. In our future work, we plan to explore the effectiveness of our proposed \textit{TGA} with this countermeasure in place.


\section {Simulation results and discussions}\label{sec:result-and-discussion}
\vspace{-5px}
In this section, we present the results and evaluate the performance of our proposed topology-guided attack on different logic locking schemes. We provide an in-depth analysis for key prediction accuracy of the proposed attack on ISCAS’85~\cite{bryan1985iscas} and ITC’99~\cite{davidson1999itc} benchmark circuits locked with \textit{RLL} and \textit{SLL} using our in-house script. In addition, we have validated our proposed attack on TrustHub benchmark circuits~\cite{TrustHub}. 

\vspace{-10px}
\subsection{Performance Analysis}
\vspace{-5px}
Four different benchmark circuits, \textit{c6288}, \textit{c5315}, \textit{b15}, \textit{b17} are first selected for determining the success rate (\textit{SR}) and misprediction rate (\textit{MR}) of our proposed \textit{TGA}. We have created 100 instances of the locked circuit based on \textit{RLL} and \textit{SLL}) for each benchmark circuits, where 128 key gates are placed, and then attacked using Algorithm~\ref{alg:prediction}. For each locked circuit, the success rate (\textit{SR}) is computed using Equation~\ref{eqn:SR}, while the misprediction rate \textit{MR} is calculated using Equation~\ref{eqn:misprediction}. In general, the mean and 
standard deviation are presented by $\mu$ and $\sigma $ for Gaussian distributions related to \textit{SR}, whereas they are all represented by $\lambda ^{-1}$  for exponential distributions that is related to \textit{MR} plots.

\begin{figure*}[!ht]
\centering
\includegraphics[width=1\linewidth]{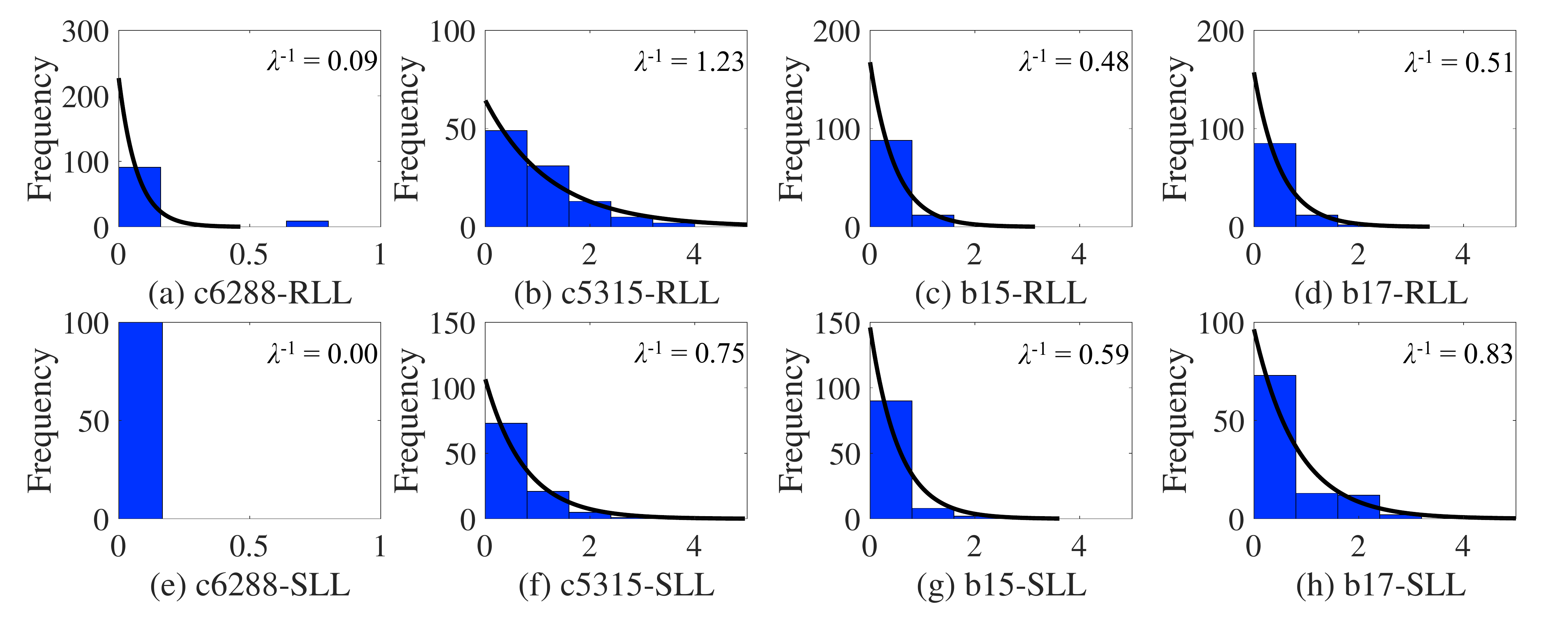} \vspace{-20px}
\caption{Histogram plots of the \textit{MR} for different \textit{RLL} and \textit{SLL} benchmark circuits with 128 key bits: (a) \textit{c6288-RLL} (b) \textit{c5315-RLL} (c) \textit{b15-RLL} (d) \textit{b17-RLL} (e) \textit{c6288-SLL} (f) \textit{c5315-SLL} (g) \textit{b15-SLL} (h) \textit{b17-SLL}} 
\label{fig:MispredictionR}
\end{figure*}

Figure~\ref{fig:SuccessRate} shows the histogram plots of \textit{SR} metric for the four selected benchmark circuits based on \textit{RLL} (see Figure~\ref{fig:SuccessRate}.(a)-(d)) and \textit{SLL} (shown in Figure~\ref{fig:SuccessRate}.(e)-(h)). For benchmark circuit \textit{c6288-RLL}, we estimate the majority of the key bits (Figure~\ref{fig:SuccessRate}.(a)) as this multiplier consists of many half and full adders. 127 out of 128 key bits can be predicted successfully, which results in a minimum \textit{SR} of 99.22\%. Figure~\ref{fig:SuccessRate}.(b) shows the \textit{SR} distribution for \textit{c5315-RLL} circuit. A Gaussian distribution is observed with $\mu$ of 88.56\% and $\sigma$ of 2.59. Similar behavior is observed for the other two benchmarks circuits as shown in Figure~\ref{fig:SuccessRate}.(c)  and Figure~\ref{fig:SuccessRate}.(d). The $\mu$ for $b15-RLL$ and $b17-RLL$ are 96.48\% and 96.41\% with the $\sigma$ values as 1.81 and 1.55 respectively. We observe a similar Gaussian distributions for the \textit{SR} on locked circuits using SLL~(see Figure~\ref{fig:SuccessRate}.(e)-(h)). 

The histogram plots of misprediction~(\textit{MR}) for the same selected benchmark circuits are presented in Figure~\ref{fig:MispredictionR}. Figures~\ref{fig:MispredictionR}.(a)--(d) present the \textit{MR} plot for the circuits locked with \textit{RLL}. For \textit{c6288-RLL} benchmark circuit, all the key bits can be determined correctly with a 0\% \textit{MR} in majority of the cases. The worst case is one bit misprediction, resulting in maximum value of \textit{MR} within 1\%. As for \textit{c5315-RLL}, we observe an exponential distribution with a mean and standard deviation ($\lambda ^{-1}$) of 1.23. As observed from Figure~\ref{fig:MispredictionR}.(c) and Figure~\ref{fig:MispredictionR}, \textit{b15-RLL} shows $\lambda ^{-1}$ of 0.48\%, whereas \textit{b17-RLL} shows $\lambda ^{-1}$ of 0.51. Likewise, a similar analysis can be done for \textit{MR} for the same selected benchmark circuits locked with \textit{SLL} plotted in Figure~\ref{fig:MispredictionR}.(e)--(h). 

\begin{table*}[ht]
\centering
\caption{Success rate (\textit{SR}) and misprediction rate (\textit{MR}) for estimating keys for RLL and SLL circuits.}\label{table:prediction3} \vspace{-5px}
\begin{tabular}{|M{1.3cm}|M{1.3cm}|M{1.3cm}|M{1.3cm}|M{1.3cm}|M{1.3cm}|M{1.3cm}|M{1.3cm}|M{1.3cm}|}
\hline
\multicolumn{1}{|l|}{\multirow{3}{*}{\textbf{Benchmark}}} & \multirow{3}{*}{\textbf{\begin{tabular}[c]{@{}c@{}}\# Total \\ Gates\end{tabular}}} & \multirow{3}{*}{\textbf{\begin{tabular}[c]{@{}c@{}}\# Key     \\ Gates\end{tabular}}} & \multicolumn{3}{c|}{\textbf{RLL}}                        & \multicolumn{3}{c|}{\textbf{SLL}}          \\ \cline{4-9} 
\multicolumn{1}{|l|}{}  &&  & \multicolumn{2}{c|}{\textbf{SR (\%)}} & \textbf{MR (\%)} & \multicolumn{2}{c|}{\textbf{SR (\%)}} & \textbf{(MR) (\%)}    \\ \cline{4-9} 
\multicolumn{1}{|l|}{}  &&  & \textbf{$\mu$}         & \textbf{$\sigma$}         & \textbf{$\lambda^{-1}$}        & \textbf{$\mu$} &\textbf{$\sigma$}  &  \textbf{$\lambda^{-1}$}    \\ \hline
c880   & 404   & 32      & 75.03 & 6.14  & 1.53 & 74.63  & 5.89 & 1.59   \\ \hline
c1350  & 593   & 32      & 69.25 & 5.97  & 0.00 & 69.10  & 6.41 & 0.00   \\ \hline
c1908  & 768   & 32      & 74.13 & 4.08  & 1.44 & 73.66  & 4.53 & 1.72    \\ \hline
c2670  & 1193  & 32      & 75.22 & 4.98  & 1.13 & 74.78  & 5.07 & 1.06   \\ \hline
c3540  & 1669  & 128     & 80.39 & 3.72  & 1.76 & 80.56  & 4.63 & 2.01    \\ \hline
c5315  & 2307  & 128     & 88.56 & 2.59  & 1.23 & 89.57  & 3.33 & 0.75    \\ \hline
c6288  & 2406  & 128     & 99.38 & 0.31  & 0.09 & 99.25  & 0.89 & 0.00   \\ \hline
c7552  & 3512  & 128     & 91.08 & 3.17  & 2.03 & 90.21  & 2.79 & 0.97    \\ \hline
b14    & 3461  & 128     & 94.16 & 2.00  & 0.52 & 93.67  & 2.10 & 0.92   \\ \hline
b15    & 6931  & 128     & 96.48 & 1.81  & 0.48 & 96.19  & 1.37 & 0.59    \\ \hline
b20    & 7741  & 128     & 97.17 & 1.44  & 0.25 & 96.95  & 1.48 & 0.84    \\ \hline
b21    & 7931  & 128     & 95.40 & 1.71  & 0.35 & 94.50  & 1.86 & 0.63    \\ \hline
b22    & 12128 & 128     & 96.34 & 1.25  & 0.37 & 95.78  & 1.62 & 0.77    \\ \hline
b17    & 21191 & 128     & 96.41 & 1.55  & 0.51 & 95.46  & 1.32 & 0.83    \\ \hline
b18    & 49293 & 128     & 90.25 & 2.54  & 0.29 & 89.36  & 2.96 & 0.80   \\ \hline
b19    & 98726 & 128     & 89.56 & 3.06  & 0.45 & 88.11  & 3.55 & 0.95    \\ \hline
\end{tabular} 
\end{table*}

Table~\ref{table:prediction3} shows the success rate (\textit{SR}) and misprediction rate (\textit{MR}) of our proposed  attack on different ISCAS'85~\cite{bryan1985iscas} and ITC'99~\cite{davidson1999itc} benchmark circuits locked with \textit{RLL} and \textit{SLL} techniques. The number of logic gates in the circuit and inserted key gates are presented in Columns 2 and 3, respectively. The number of key gates is set to 32 for the first four benchmark circuits (\eg c880, c1350, c1908, and c2670), while the remaining are inserted with 128 key gates. Columns 4 and 5 presents the mean value $\mu$ and standard deviation $\sigma$ of \textit{SR} values (see Equation~\ref{eqn:SR}) by analyzing 100 locked instances for each benchmark circuit to determine the accuracy of the proposed \textit{TGA} (see Algorithm~\ref{alg:prediction}  for details) for \textit{RLL}. For \textit{c5315} benchmark, 128 key gates are inserted randomly in the netlist with 2307 logic gates. The $\mu$ of success rate \textit{SR} is 88.56\%, and the $\sigma$ is 2.59\%, which means that the confidence of 99.7\% (for $\pm$3$\sigma$) can be observed in the range from  80.79\% to 96.33\%. A similar analysis can be performed for all the benchmarks shown in each row. For the larger benchmark circuits, the average success rate \textit{SR} can be increased over 90\% because of the increased search space, which makes our proposed \textit{TGA} efficient for larger designs. Note that, although \textit{SAT} fails on benchmark c6288, our proposed attack provides better accuracy (average of 99.38\%) for benchmark \textit{c6288} due to its special topology -- it is a multiplier, which consists of 225 full adders and 15 half adders. Therefore, an adversary can choose our proposed attack as an alternative to the SAT attacks.

Column 6 shows the mean (or standards deviation) \textit{MR} of each benchmark circuit, calculated using Equation~\ref{eqn:misprediction}. Note that the mean and standards deviation are of the same value for an exponential distribution. The average \textit{MR} is less than 1\% for most benchmark circuits, which makes our attack very useful for determining the secret key. Table~\ref{table:prediction3} also presents the \textit{SR} and \textit{MR} metrics for the same benchmark circuits locked with \textit{SLL}, and shown in Columns 7 to 9. We also observe a similar trend like \textit{RLL}.



\begin{figure*}[ht]
\centering
\includegraphics[width=1\linewidth]{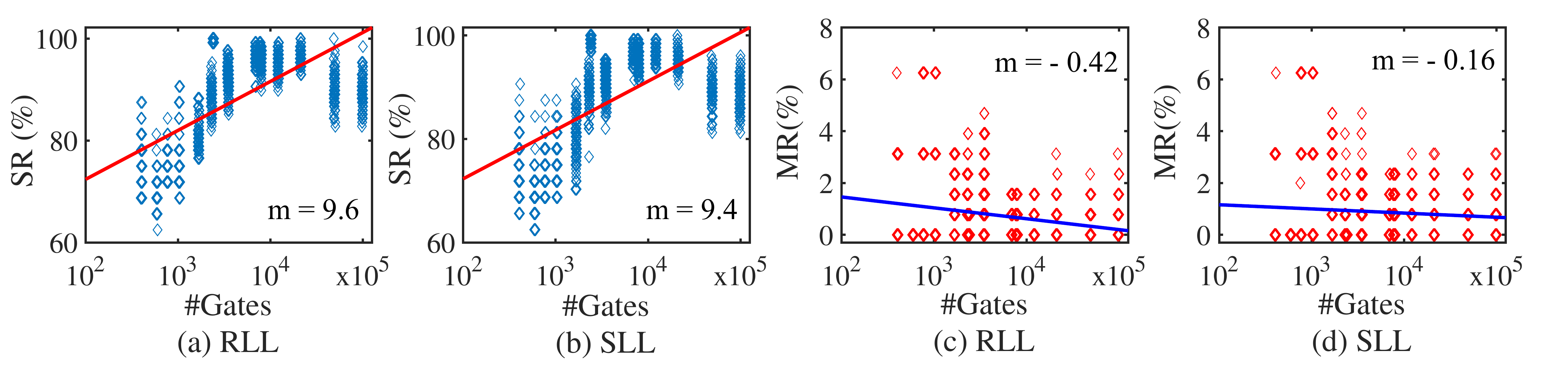} \vspace{-13px}
\caption{Scatter plots of  \textit{SR} and \textit{MR} versus number of gates on \textit{RLL} and \textit{SLL} benchmark circuits. (a) \textit{SR} for \textit{RLL} circuits, (b) \textit{SR} for \textit{SLL} circuits, (c) \textit{MR} for \textit{RLL} circuits, and (d) \textit{MR} for \textit{SLL} circuits.  } \label{fig:Scatter}
\end{figure*}

In order to evaluate the accuracy of \textit{SR} and \textit{MR} with the circuit size, scatter plots of \textit{SR} and \textit{MR} are performed and shown in Figure~\ref{fig:Scatter}. A least-squares trend line is added in every plot, while $m$ represents the line's slope.  In Figure~\ref{fig:Scatter}.(a), we observe a slope $m$ of 9.60, which indicates that the value of \textit{SR} increases with the circuit size increase. A similar trend is observed for the \textit{SLL} locked circuits, and shown in Figure~\ref{fig:Scatter}.(b). As for the \textit{MR} shown in Figure~\ref{fig:Scatter}.(c), we observe a negative slope of 0.42. A similar trend is found for \textit{SLL} locked circuits and shown in Figure~\ref{fig:Scatter}.(d). From this observation, we can conclude that the misprediction rate decreases with the increase of the circuit size. Overall, the accuracy of the proposed \textit{TGA} increases for larger circuits.

\begin{table}[!ht]
\centering
\vspace{-10px}
\caption{\textit{SR} and \textit{MR} for estimating keys for locked circuits from Trust-Hub.} \label{table:SLL} \vspace{-5px}
\begin{tabular}{|M{1.8cm}|M{1.1cm}|M{1.1cm}|M{1cm}|M{1cm}|} 
\hline
Benchmark    & \# Total Gates & \# Key Inputs & \multicolumn{1}{c|}{\textbf{\textit{SR}(\%)}} & \multicolumn{1}{c|}{\textbf{\textit{MR}(\%)}} \\ \hline
c880-SL320   & 404   & 32   & 87.50   &  3.13   \\ \hline
c1350-SL320  & 593   & 32   & 78.13  & 0.00   \\ \hline
c1908-SL320  & 768   & 32   & 84.38   &  3.13 \\ \hline
c2670-SL320  & 1042  & 32   & 84.38   &  3.13 \\ \hline
c3540-SL640  & 1546  & 64   & 82.81   &  1.56 \\ \hline
c5315-SL640  & 2090  & 64   &  87.50  &  1.56\\ \hline
c6288-SL1280 & 2603  & 128  & 96.88   &  0.00  \\ \hline
c7552-SL1280 & 3173  & 128  & 88.28   &  0.78  \\ \hline

\end{tabular}
\end{table}

To reinforce our conclusion from Table~\ref{table:prediction3}, we also selected 8 different benchmark circuits from trust-Hub~\cite{TrustHub} and performed our topology-guided attack to evaluate the effectiveness. Table~\ref{table:SLL} presents the obtained results for the same. The selected benchmark is noted in Column 1 with the corresponding number of logic gates in the circuit shown in Column 2. Columns 3 presents the number of key inputs instead of key gates for each benchmark circuits as one key input may be fed into multiple key gates. The resultant \textit{SR} and \textit{MR} are concluded in Columns 4 and 5. For c3540-SL640, 64 key inputs are inserted in the circuit. The \textit{SR} is 82.81\%, which depicts 53 key inputs can be predicted with correct values. When comparing the results, 1 incorrect prediction is found, which produces a  misprediction rate \textit{MR} of 1.56\%. As for the c6288-SL1280 benchmark circuit, 128 key inputs are inserted to protect the circuit. The \textit{SR} of 96.88\% can be observed which indicates that the majority of key inputs can be recovered based on our attack (125 key bits). The \textit{MR} is 0.00\%, which 0 key bit is mispredicted out of the entire 128 key inputs. We have emphasized \textit{c6288} benchmark circuit as to present a clear comparison with the SAT attack, which was not efficient on this circuit.

Note that an adversary can recover the complete key from Table~\ref{table:prediction3} with the help of an oracle (e.g., an unlocked chip). As the objective of logic locking is to modify the input-output response of a circuit, it produces incorrect responses for applying a wrong key. If an adversary finds out the key bit location (the unspecified, $X$, key bits from Algorithm~\ref{alg:dfs}, it is easy to determine its value by comparing it with an oracle). As there are few $X$s, their permutations are limited and can easily be determined. Note that this could have a complex problem if we do not know the location of the wrong key bit(s). Then, the adversary needs to verify $^{|K|}C_N \times 2^{N}$, where $N$ is the number of unspecified keys, and $|K|$ is the key size. These $N$ unspecified key bits can come from any locations of the key $K$. On the contrary, we only need to verify $2^{N}$ cases to determine the complete key, which is a much simpler problem.

\subsection{Complexity Analysis}
SAT problem is an NP-complete problem, thus solving an SAT-resistant locking leads to an exponential worst-case complexity. 
However, our proposed topology-guided attack does not need to compare any input and output pairs, and all the inserted key gates are analyzed individually. Therefore, the time complexity of the attack itself is simply linear to the key size, namely, $O(|K|)$. Note that, our attack algorithm is based on \textit{FS}, the actual overall complexity is $O(|K|*n*u)$ where $n$ and $u$ represent the size of the netlist and maximum size of the unit functions, respectively. Thus, the complexity could be considered as linear for a particular circuit, since the netlist size is fixed, and the size of \textit{UF} normally ranges from 3-10 gates, depending on the key gate location. In Algorithm~\ref{alg:prediction}, once a key bit is predicted and written in the key list $K_P$, it will never be analyzed again as the value is recovered already. As a result, the computation complexity of launching the attack on \textit{SLL} is the same as it is for \textit{RLL}. 

\begin{figure}[!ht]
\centering
\vspace{-10px}
\includegraphics[width=1.0\linewidth]{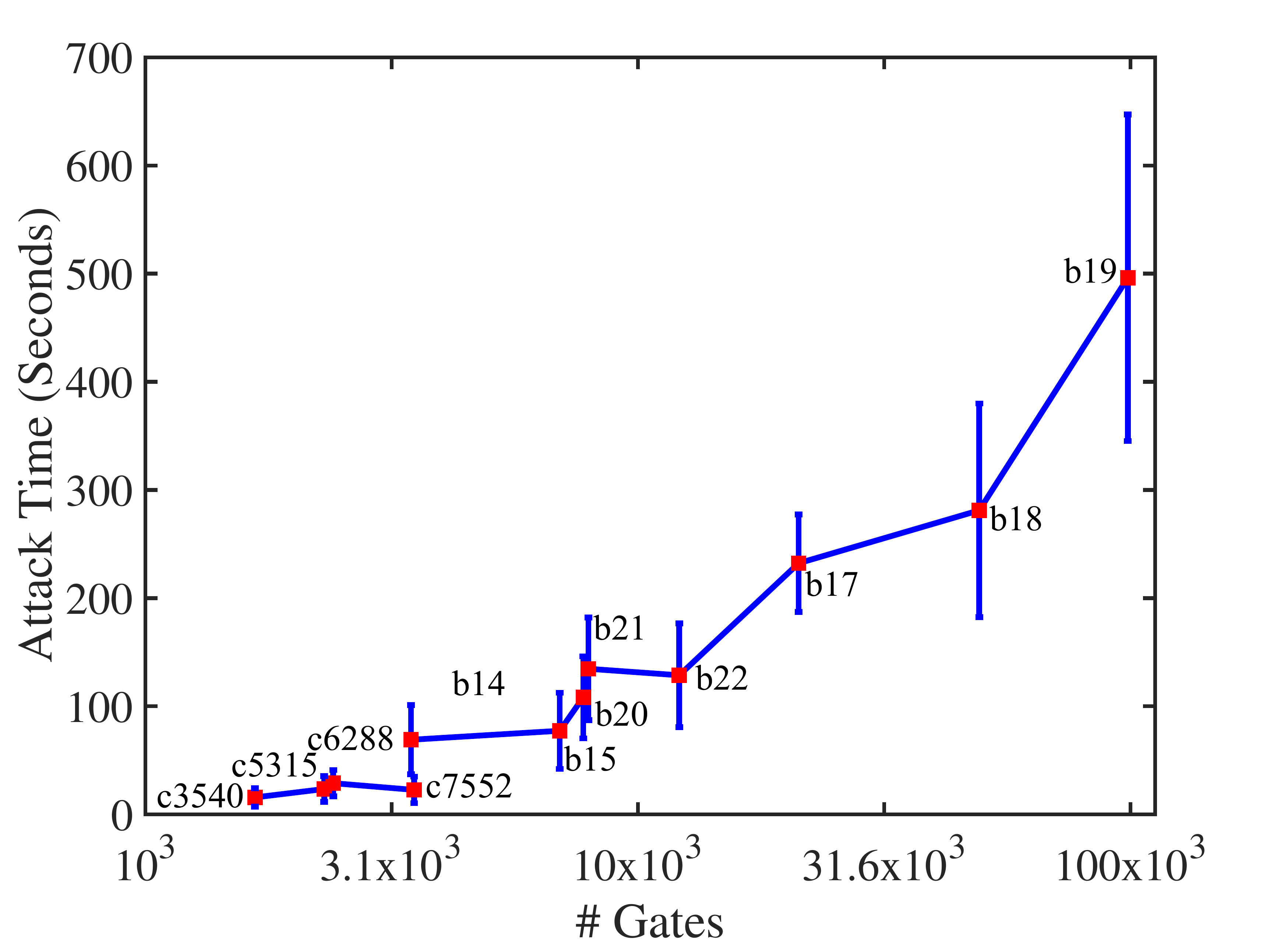} \vspace{-20px}
\caption{Attack time for different \textit{SLL} locked benchmark circuits with 128-bit keys . Each benchmark circuit is evaluated with 100 instances.} 
\label{fig:attack_time}
\end{figure}

The attack time of the proposed \textit{TGA} on different \textit{SLL} locked circuits is illustrated in Figure 9. The plot shows the attack time versus the number gates for the 100 instances of each benchmark circuit (same circuits mentioned in Table 1). Note that we consider benchmark circuits inserted with 128 key gates only for the uniformity. The minimum, average, and maximum values of the attack time are displayed for each benchmark circuit. For example, the reported time for performing \textit{TGA} on 128-bit \textit{SLL} locked b17 benchmark circuit lies in a range from 178 seconds to 277 seconds with an average of 222.38 seconds. The location of inserted key gates causes this variation, and the corresponding number of \textit{EUFs} searched during the attack. The x-axis is presented in the log scale to display all the different benchmarks clearly. The graph displays exponential behavior, and thus, it can be concluded that the attack time increases linearly with the increase of circuit size.

\section {Conclusion}\label{sec:conclusion}
In this paper, we proposed a novel oracle-less topology-guided attack that is based on unit function search. Due to the repetitive usage of \textit{UF} in a netlist, the key bits for a locked unit functions can be determined by constructing \textit{EUFs} with hypothesis key bits and comparing them against the corresponding unlocked \textit{UFs}. Compared to the traditional SAT-based attacks, the proposed topology-guided attack does not require input/output pairs or an activated chip. Moreover, SAT resistant countermeasures cannot prevent an adversary from launching this attack. To demonstrate the success of this attack, we presented the results on different benchmark circuits locked with random logic locking and strong logic locking techniques. We also validated our proposed attack on existing locked benchmark circuits from the trust-Hub. The success rate and misprediction rate metrics are proposed to evaluate the effectiveness of this attack. It is important to emphasize on the complexity of this attack which is linear with the key size on both \textit{RLL} and \textit{SLL}, which makes it very effective for circuits with larger key sizes. A countermeasure is also proposed as a solution to prevent this topology-guided attack. The basic idea is to insert the key gate in a unique unit function or lock all the instances repeated in the netlist. Note that this solution can only be used to prevent this topology-guided attack. To design a secure logic locking technique, one needs to select an existing secure logic locking technique along with our proposed solution. 

\section*{Acknowledgement}
This work was supported by the National Science Foundation under grant number CNS-1755733. Any opinions, findings, and conclusions or recommendations expressed in this material are those of the authors and do not necessarily reflect the views of the National Science Foundation.

\bibliographystyle{spmpsci}
\bibliography{main}

\end{document}